# Selective tracking of charge carrier dynamics in CuInS$_2$ quantum dots

Andrés Burgos-Caminal,*[1,2] Brener R. C. Vale,[3,4] André F. V. Fonseca,[5,4] Elisa P. P. Collet,[1,2] Juan F. Hidalgo,[1] Lázaro García,[2] Luke Watson,[6] Olivia Borrell-Grueiro,[7] María E. Corrales,[2,8] Tae-Kyu Choi,[9] Tetsuo Katayama,[10] Dongxiao Fan,[11] Víctor Vega-Mayoral,[1] Saül Garcia-Orrit,[1] Shunsuke Nozawa,[11] Thomas J. Penfold,[6] Juan Cabanillas-González,[1] Shin-Ichi Adachi,[11] Luis Bañares,[7,1] Ana F. Nogueira,[5] Lázaro A. Padilha,[3] Marco A. Schiavon,[4] and Wojciech Gawelda*[2,1,12]

1. Madrid Institute for Advanced Studies IMDEA Nanoscience, Ciudad Universitaria de Cantoblanco, Calle Faraday 9, 28049 Madrid, Spain.
2. Departamento de Química, Universidad Autónoma de Madrid, Ciudad Universitaria de Cantoblanco, Calle Francisco Tomás y Valiente 7, 28049 Madrid, Spain.
3. Instituto de Física Gleb Wataghin, Universidade Estadual de Campinas - UNICAMP, Campinas, São Paulo, Brazil.
4. Grupo de Pesquisa Química de Materiais, Departamento de Ciências Naturais, Universidade Federal de São João Del-Rei, Brazil.
5. Laboratório de Nanotecnologia e Energia Solar, Instituto de Química, Universidade Estadual de Campinas – UNICAMP, Campinas, São Paulo, Brazil.
6. Chemistry-School of Natural and Environmental Sciences, Newcastle University, Newcastle upon Tyne, NE1 7RU, UK.
7. Departamento de Química Física and Center for Ultrafast Lasers, Facultad de Ciencias Químicas, Universidad Complutense de Madrid, 2804 Madrid, Spain
8. Departamento de Química Física Aplicada, Universidad Autónoma de Madrid, Ciudad Universitaria de Cantoblanco, Calle Francisco Tomás y Valiente 7, 28049 Madrid, Spain.
9. XFEL Division, Pohang Accelerator Laboratory, POSTECH, Pohang, Gyeongbuk 37673, Republic of Korea.
10. Japan Synchrotron Radiation Research Institute, Kouto 1-1-1, Sayo, Hyogo 679-5198, Japan.
11. Institute of Materials Structure Science, High Energy Accelerator Research Organization (KEK), Japan.
12. Faculty of Physics, Adam Mickiewicz University, ul. Uniwersytetu Poznańskiego 2, 61-614 Poznań, Poland

Corresponding authors: wojciech.gawelda@uam.es, andres.burgos@imdea.org

## Abstract

CuInS$_2$ quantum dots have been studied in a broad range of applications, but despite this, the fine details of their charge carrier dynamics remain a subject of intense debate. Two of the most relevant points of discussion are the hole dynamics and the influence of Cu:In synthesis stoichiometry on them. It has been proposed that Cu-deficiency leads to the formation of Cu$^{2+}$, affecting the localization of holes into Cu defects. Importantly, it is precisely these confined hole states which are used to explain the interesting photoluminescence properties of CuInS$_2$ quantum dots. We use static X-ray spectroscopy to reveal no evidence for a measurable amount of native Cu$^{2+}$ states in Cu-deficient samples. Instead, the improved properties of these samples are explained by an increase of crystallinity, reducing the concentration of mid gap states. Furthermore, to understand the charge carrier dynamics, herein we employ ultrafast optical transient absorption, and fluorescence up-conversion spectroscopies in combination with ultrafast X-ray absorption spectroscopy using a hard X-ray free electron laser. We demonstrate that in non-passivated samples, holes are transferred from Cu atoms in sub-picosecond timescales. We assign this transfer to occur towards the thiol-based ligands. Finally, we observe that Cu-deficient samples are more robust against the photothermal heating effects of using higher laser fluences. This is not the case for the stoichiometric sample, where heating effects on the structure are directly observed.

**Keywords:** CuInS$_2$ Quantum Dots, XFEL, X-ray absorption, Transient Absorption, charge carrier dynamics, hole trapping.

CuInS$_2$ quantum dots (CIS QDs) have found applications across diverse fields including photovoltaics,[1–4] LEDs,[5–7] solar concentrators,[8–10] and biological labeling.[11,12] This diversity of applications comes from their unique optical properties such as: high molar absorption coefficient (> 10$^4$ L mol$^{-1}$ cm$^{-1}$), large Stokes shift (from 200 to 400 meV) which makes them free of reabsorption effects, long charge carrier recombination times (more than hundreds of nanoseconds), and fluorescence that can be tuned by the diameter of the QD.[13] Furthermore, they are free of the highly toxic Cd or Pb, typically used in other QDs. However, CIS QDs usually have a low photoluminescence quantum yield (PLQY) due to a high density of defects. In order to overcome this limitation, approaches involving core-shell structures have been used, achieving PLQY values as high as 50%.[14–16]

The most accepted model for the photoluminescence (PL) mechanism of CIS QDs is that it derives from a recombination of a delocalized electron in the conduction band with a confined hole state (CHS).[14,17–20] This CHS is related to Cu defects, which are about 100 meV above the material's valence band. Ultrafast spectroscopy measurements suggest that after the photoexcitation, hole trapping occurs on a timescale spanning from hundreds of femtoseconds to a few picoseconds.[14,17,21] This is followed by two competing processes: electron trapping at sub-band gap trap states located mainly at the QD surface,[22,23] and radiative recombination of the electron with the CHS.[14,17,21]

The oxidation state of the Cu-related defects, acting as localization centers for the CHS, also plays a critical role in controlling the optical properties of CIS QDs. With respect to absorption, the current understanding is that defects of Cu$^+$ have an active absorption fingerprint due to excitation from a 3$d^{10}$ state to the conduction band of the material, which would be responsible for the broad absorption tail below the bandgap.[21,24] Contrarily, defects related to Cu$^{2+}$, already containing a CHS, do not affect the absorption spectrum with such a tail because a transition from a 3$d^9$ electronic configuration would require energy above the bandgap. Furthermore, the recent literature explains the emission mechanism from the CHS in the following way: after photoexcitation in the material's bandgap, a fast hole transfer from the valence band to the Cu$^+$ defects must occur, forming the CHS. Alternatively, the emission from native Cu$^{2+}$ defects promptly occurs due to the already present vacancy in the 3$d^9$ electronic configuration; however, another hole trapping mechanism, not related to Cu, needs to quickly remove the photogenerated hole from the valence band to account for the observed high PLQY. Otherwise, an Auger recombination involving the two holes (valence band and $d^9$ vacant orbital) and the electron in the conduction band would suppress the emission.[21,22]

According to recent reports, an efficient way to control the defects related to Cu is by varying the QD stoichiometry during synthesis. Based on charge neutralization arguments, in stoichiometric samples (Cu/In = 1), most of the defects are cation antisite, where $Cu^+$ is located in a site of $In^{3+}$ with a double negative charge and an $In^{3+}$ is localized in a $Cu^+$ site with a double positive charge. Following the Kröger-Vink notation of crystallographic defects, this is represented as $Cu_{In}''$-$In_{Cu}^{\cdot\cdot}$. Moreover, Cu deficiency leads to Cu vacancy ($v_{Cu}'$), therefore, charge neutralization creates $Cu^{2+}$ ($Cu_{Cu}^{\cdot}$).[13,21,22] In addition, CIS QDs have low PLQY due to surface trap states and, to increase this parameter, the QDs are shelled with a wide bandgap semiconductor material. One good candidate for this task is the insulator ZnS.[12,25,26]

Typically, the characterization of CIS QDs, and especially of their charge carrier dynamics, is done employing UV-Vis and sometimes NIR spectroscopy techniques.[13,17,22,27] Some studies have included the use of X-ray absorption spectroscopy (XAS),[28] including time-resolved XAS (TR-XAS).[29] However, no study before has been able to take advantage of TR-XAS with femtosecond time resolution, available thanks to the recent development of X-ray free electron lasers (XFELs). Measuring physicochemical properties in colloidal suspension is advantageous because procedures for obtaining nanocrystals (NCs) as a solid can affect their properties. Based on that, UV-Vis absorption and PL measurements have been the most used techniques to characterize different processes due to their simplicity in their application to colloids. However, other more specific approaches that allow the characterization of NCs in a colloidal dispersion are highly desirable. Here, we used steady-state XAS and femtosecond TR-XAS to investigate colloidal CIS QDs. X-ray spectroscopy has a considerable advantage with respect to UV-Vis. It is sensitive to structural and oxidation state changes, and it has element specificity. This means that we can probe the oxidation state changes in Cu upon the formation of a valence band hole, or the structural changes between samples.[30] In addition, using XAS techniques enables us to measure structural properties directly in a colloidal suspension, unlike other X-ray or electron techniques such as X-ray powder diffraction, X-ray photoelectron spectroscopy, and transmission electron microscopy. Here, we have employed, for the first time, femtosecond hard X-ray pulses from the SACLA XFEL facility (Hyogo, Japan) to study the ultrafast dynamics of CIS QDs. We have studied three different samples to understand how stoichiometry and the presence of a passivation layer affects the optical properties of the material. Our main findings are concentrated in understanding the role of Cu:In stoichiometry and surface passivation on the behavior of both charge carriers. We probed the effects on the steady-state UV-Vis and X-ray spectra, and the corresponding ultrafast dynamics of CIS QDs, showing results that contradict some of the current understanding of the photophysics and the compositional effects, prompting further investigation into these systems.

## Results and discussion

We investigated three different CIS QD samples in order to understand the effects of stoichiometry and surface passivation on the structure and charge carrier dynamics: a) stoichiometric bare CIS QDs (CIS100%), b) Cu-deficient bare CIS QDs (CIS20%), and c) a passivated Cu-deficient sample, doped with Zn and submitted to a post-treatment with $ZnCl_2$ at 200ºC (CZIS200ºC). This treatment gives the QDs a core-shell structure, passivating the surface and minimizing non-radiative recombination. Therefore, this last sample allows us to better observe the hole behavior without the interference of other trap-assisted processes at the surface. The detailed synthesis method, based on previously published work,[14,31] can be found in the methods section, yielding CIS QDs expected to have chalcopyrite structure[32–34] (Fig 1.A), although it may be difficult to distinguish from a face-centered cubic structure.[31] The UV-Vis absorption and PL (Fig. 1.B) serve as a first characterization of the samples. We can already identify prominent Urbach tails below the excitonic absorption energy at 550 nm in the non-passivated samples CIS100% and CIS20%. Furthermore, this tail is considerably larger in CIS100% and the excitonic band is not distinguishable, pointing to a higher degree of disorder. Last, only the well-passivated CZIS200 ºC sample shows a considerable PL signal (Fig. S1), which correlates with the lower density of trap states producing an Urbach tail and having a PLQY of 29.8 ± 0.3 %. These changes in absorption and PL give clear differences in the appearance of the samples (Fig 1.C).

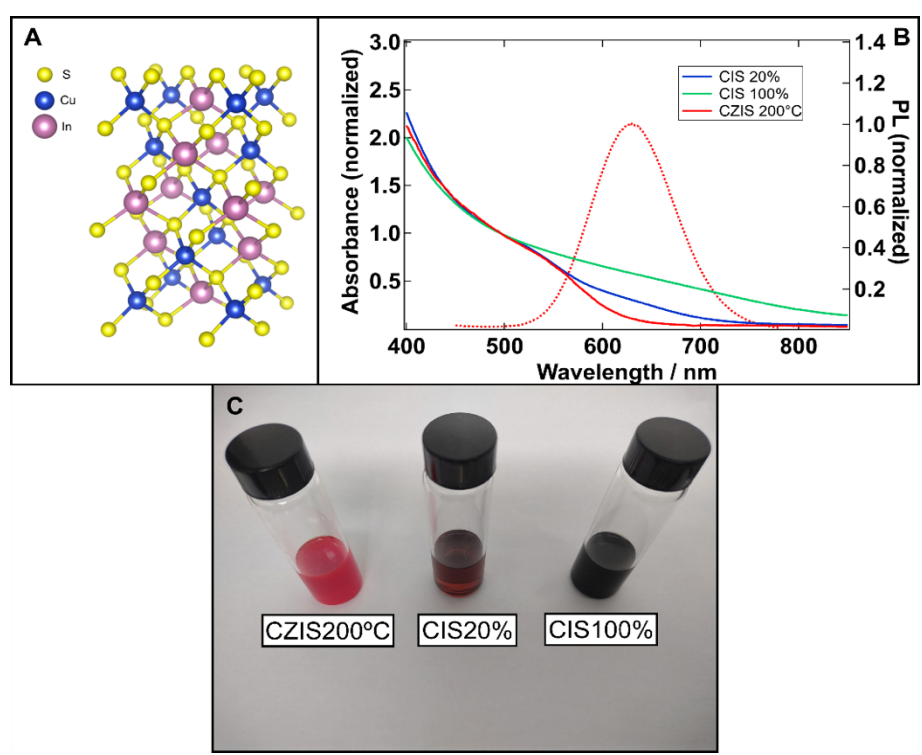

**Figure 1**. A) Crystal structure of chalcopyrite $CuInS_2$.[32,35] B) Steady-state absorption spectra of the three samples under study (solid lines, normalized at 500 nm), as well as the PL signal of CZIS200°C, the only highly emissive sample (dashed line). C) The three samples under study: CZIS200ºC, CIS20% and CIS100%.

## Cu K-edge X-ray absorption spectroscopy

We measured the Cu K-edge X-ray absorption spectra of our three samples at the CLÆSS beamline of the ALBA synchrotron, with the goal of studying the effect of stoichiometry and finding evidence for the presence of $Cu^{2+}$ in Cu-deficient samples (Fig. 2). We obtained both the X-ray absorption near-edge structure (XANES) and the extended X-ray absorption fine structure (EXAFS) spectra.

At first glance, in Fig. 2A, it may appear that the K-edge of CIS100% is shifted to lower energies compared to the others. However, as it will be explained below, this is not true. In addition, the stoichiometric sample shows a suppressed white line at 8984 eV compared to the non-stoichiometric CIS20% and CZIS200ºC. This band is ascribed to the 1s→$4p_z$ transition, and its energy position and intensities are sensitive to the local structure of the $Cu^+$ species.[36] Simulations reported by Stan et al.[28] show that a more intense transition is associated with an increase of positive charge on the absorbing atom, and thus implies an increase in its average oxidation state. Such behavior was studied experimentally by applying electrochemical potentials, but the changes in the XANES spectrum were not conclusive.[28] Hu et al. have demonstrated that this peak is enhanced with increasing NCs size.[29] Furthermore, this enhancement is attributed to tetrahedrally coordinated $Cu^+$ and increasing uniformity, while a smoother feature is characteristic of a triangular coordination and disorder.[29,37] Therefore, the changes observed in our data can either be due to a decrease in the disorder of the tetrahedral structure or to a change in the oxidation state.

A typical signature that reveals the presence of $Cu^{2+}$ is the presence of a pre-edge peak in the 8975-8980 eV region due to a quadrupole-allowed 1s→3d transition.[37,38] An example can be seen in a reference spectrum in Fig. S2. However, all three CIS QDs show no such feature in their XANES spectra or their derivative (Fig. 2A and B). Thus, we exclude the possibility of considerable changes in the oxidation state. Furthermore, no consistent shift is observed in the different features, compared to the $Cu^+$ reference sample (Fig. S3). Indeed, the post-edge peak at 8995 eV, ascribed to the 1s→$4p_{xy}$ transition of $Cu^+$, is intensified while moving from stoichiometric to non-stoichiometric samples, but does not shift. Moreover, spectroelectrochemical measurements show that the PLQY of QDs decreases upon applying an oxidative potential through the formation of $Cu^{2+}$.[28] However, our PL data shows the opposite behavior: the PL is enhanced for non-stoichiometric samples, which are supposed to have higher density of $Cu^{2+}$.[22] Therefore, Cu K-edge XANES measurements suggest that non-stoichiometric samples favor the tetrahedral coordination of Cu-S and a decrease in disorder. This conclusion is supported by other studies in the literature.[29,37,39] Furthermore, we can simulate the same behavior of the XANES spectra by introducing interstitial defects as a representation of disorder (Fig. 2C).

In Fig. 2D we present the Cu K-edge EXAFS. We note that the main feature centered at ~1.9 Å has higher intensity for the non-stoichiometric CIS20% and CZIS200ºC. This is associated with either an increase in the coordination number or a decrease in disorder around the metal center.[29] We performed a FEFF6 fitting analysis of the Cu K-edge EXAFS data with the Demeter package,[32,40] and we observed that the coordination number of Cu increases from 2 for the stoichiometric sample to 3 for the Cu-deficient ones, pointing to a larger disorder and defects in the former. Certainly, the ideal value is expected to be higher for the bulk material, since each Cu atom is surrounded by 4 S atoms. However, coordination numbers between 2 and 3 agree with the data from the literature for CIS QDs.[29,37] The reduced coordination number is due to the large proportion of surface atoms compared to the total volume. Moreover, the fitted bond length of Cu-S increases slightly while moving from stoichiometric to non-stoichiometric samples. These results are summarized in Table S2, and the corresponding fit curves are shown in Fig. S4. In short, Cu K-edge EXAFS results confirm that the Cu deficiency increases the coordination number and decreases disorder. It is worth noting that these results can also be explained by a transition from an In-dominated surface in Cu-deficient CIS20% and CZIS200ºC, to a Cu-dominated one for the stoichiometric CIS100%. This would induce a general decrease in Cu coordination and may explain the introduction of defect states that serve as recombination centers.

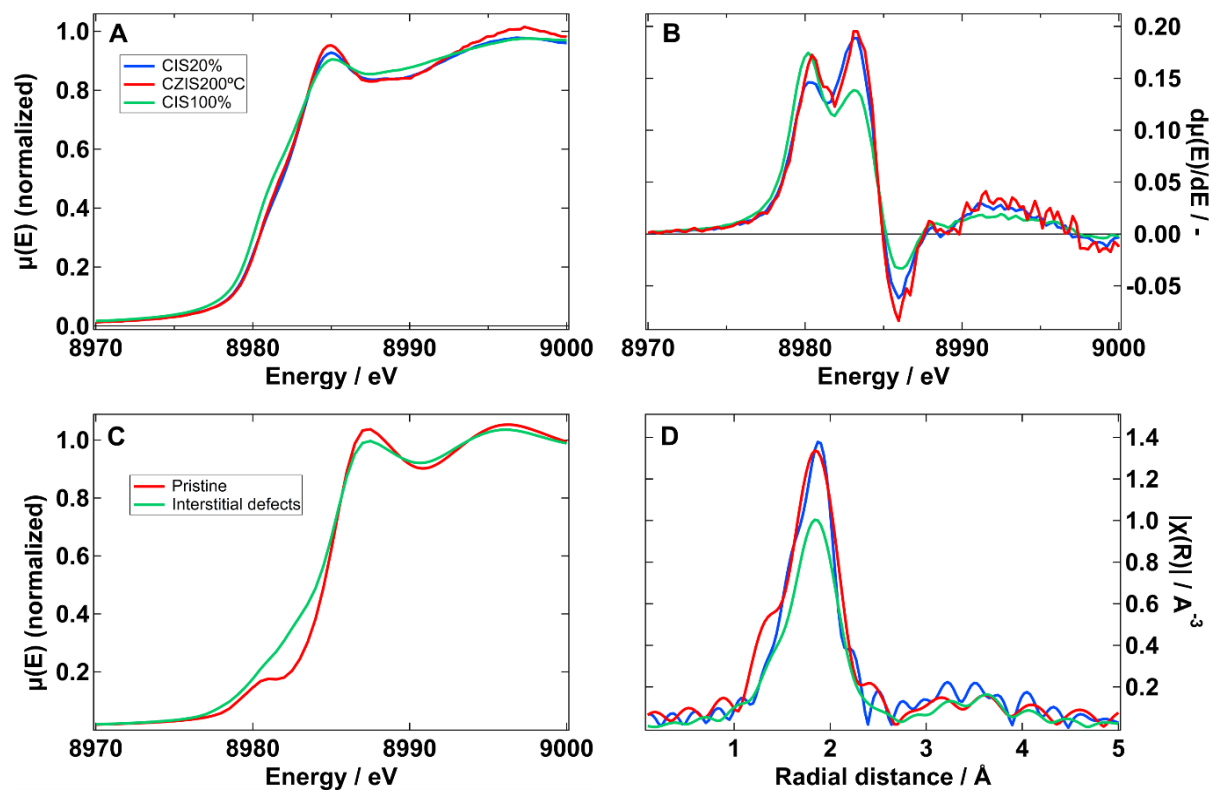

**Figure 1**. A) XANES spectra of the three samples under study. B) First derivative of the XANES spectra lacking any evidence of a pre-peak around 8978 eV. C) Simulations of XANES spectra including interstitial defects to simulate disorder. D) EXAFS of the corresponding samples. The main differences between the stoichiometric and non-stoichiometric samples are a broadening of the XANES peaks, and a lowering of the EXAFS one. This is attributed to disorder in the sample. A similar effect is seen in the simulations. A, B, and D share the same legend.

**Transient absorption spectroscopy**

We now turn to time-resolved measurements to study the effect of composition on the charge carrier dynamics. First, we analyze ultrafast optical transient absorption (OTAS) measurements, which are mainly sensitive to the population of charge carriers, in the form of a strong ground state bleach (GSB) and weak excited state absorption (ESA). This is seen as a negative ΔA signal in a broad region, resonant with the excitonic absorption of the sample, and a positive signal, respectively.[41,42] Due to the much larger effective mass of holes in these systems, OTAS is mainly sensitive to the evolution of electrons.[13,43]

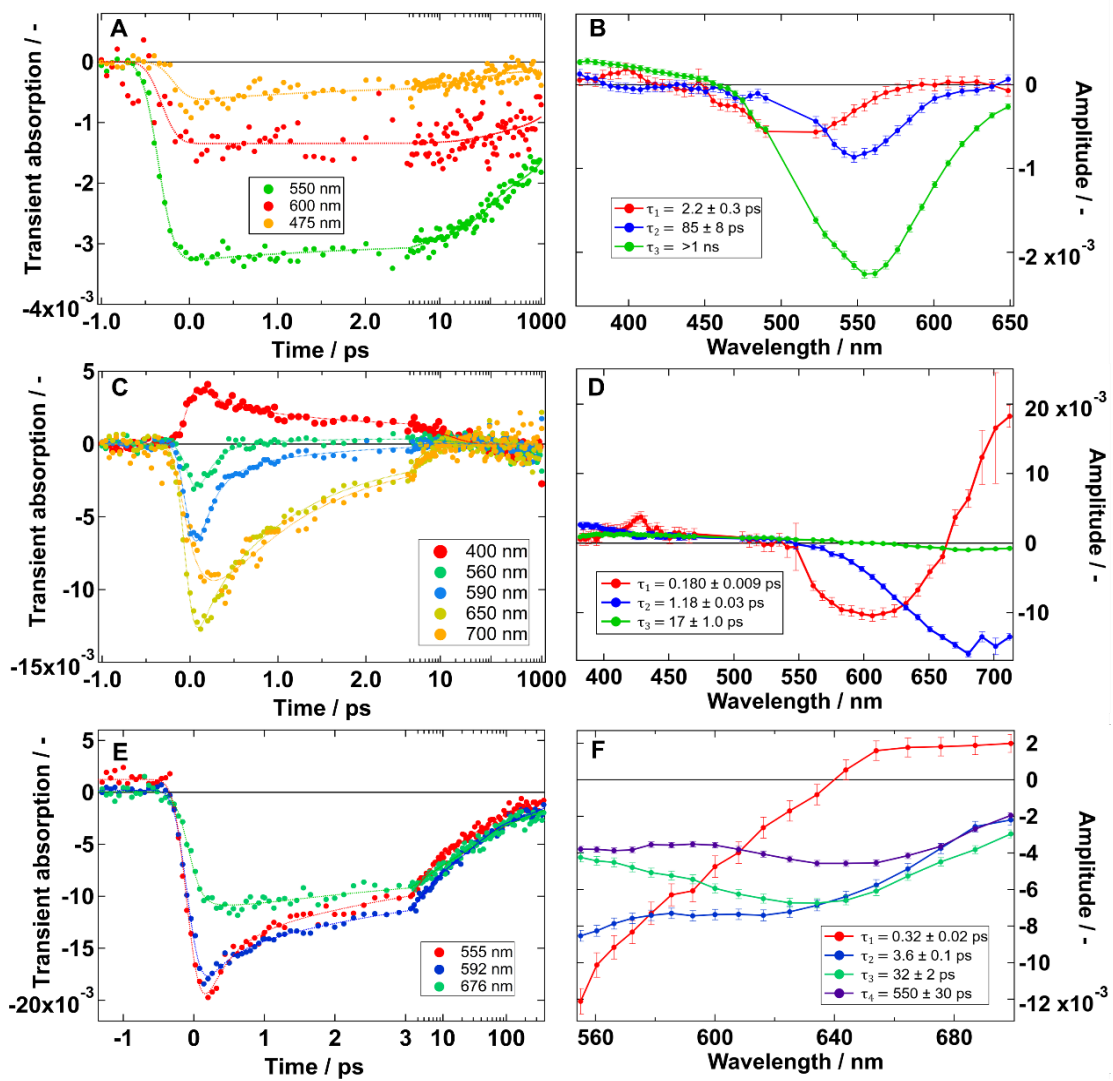

**Figure 2.** OTAS of CZIS200°C (A, B), CIS100% (C, D), and CIS20% (E, F) pumped at 520 nm and 0.08, 0.6 and 1.0 mJ·cm$^{-2}$, respectively. The left shows the time evolution at selected wavelengths while the right shows the decay-associated spectra obtained through a global fit to a multi-exponential model. We used low fluence in CZIS200°C to avoid multi-excitonic effects. In contrast, we used higher fluences for the other two samples because: a) trapping dominates over multi-excitonic effects, as seen on the X-ray measurements below, and b) it was needed to improve the signal to noise in these two systems.

In Fig. 3 we show the OTAS results for our three distinct samples, CZIS200°C, CIS100%, and CIS20%. We can observe that the probed dynamics at different wavelengths of CZIS200ºC are almost flat in a time window of 4 ps, but for CIS100% the GSB decays much faster. Alternatively, the behavior of CIS20% lies in between the other two. We apply a global fit of a three- or four-exponential model to characterize the time evolution of the transient absorption signal. This analysis results in the corresponding fit lines on the left plots and the decay-associated spectra (DAS) in the right plots. The DAS shows the wavelength dependence of the amplitude of each of the fitted lifetimes. Thus, it allows us to interpret spectral differences between decays. A representation of the different charge carrier evolution pathways is shown in Scheme 1 and referenced throughout the text with numbers **1** to **6**. We observe a blueshift in the main GSB signal when going progressively from CIS100% to CIS20% and to CZIS200°C owing to the effects of Zn-doping and stoichiometry, similar to the effects on steady-state absorption (Fig. 1). Interestingly, the shortest decay shows a different DAS from the other ones. It is seen as an early decay of the blue side of the GSB on CZIS200°C and as a shift of the GSB in CIS100% and CIS20%. Previous reports have attributed these changes in the GSB to hole trapping into the CHS (**1**).[14,44,45] This may be true for the relatively small effect of $\tau_1$ on the CZIS200°C sample. However, the large effects seen on the bare samples must be related to fast charge trapping at states below the excitonic band (**2** and **3**), probably related to the surface. Cooling of charge carriers in the bands is not observed since we pump at the excitonic band with 520 nm. Later lifetimes correspond to the recombination of the trapped carriers (**4**). This explains the lack of PL in CIS100% and CIS20%. In addition, it confirms that the tails of absorption below the excitonic peak in both samples correspond to abundant trap states related to 1) the bare surface (for both), and 2) the disorder in CIS100%. The results also revealed that the non-stoichiometric sample recovered much slower than the stoichiometric one due to the lack of the second source of traps from internal defects. In the case of the passivated sample, CZIS200ºC, we can ascribe the second (A2) and third (A3) time constants to electron trapping (**3**) and nonradiative recombination (**5**), in which trapped electrons recombine with the hole in the CHS. These lifetimes correspond to a portion of the QDs that are susceptible to trapping due to the lower presence of defects compared to the other samples.[14,17] From time-resolved PL (Fig. S9 and table S3) measurements of the most passivated sample (CZIS200°C), we can deduce a radiative recombination (**6**) lifetime of *ca.* 300 ns.

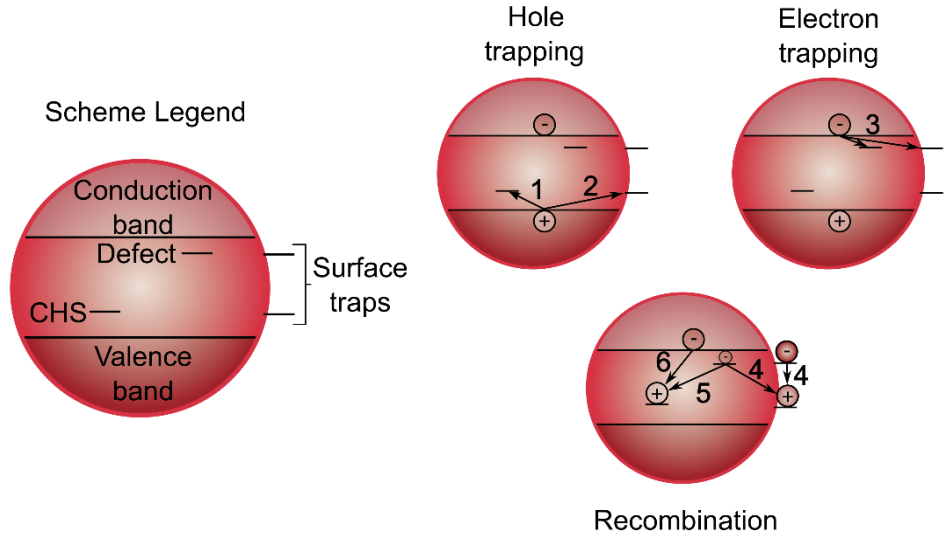

**Scheme 1**. Representation of charge carrier pathways. 1) CHS formation, 2) hole surface trapping, 3) electron trapping, 4) Nonradiative recombination of trapped carriers, 5) nonradiative recombination with CHS, and 6) radiative recombination with CHS.

In agreement with the steady-state data, the non-stoichiometric samples present lower electron traps and nonradiative rates (Table 2), confirming that those samples have much better optical properties than the stoichiometric ones.

**Table 1.** Electron trapping (**3**) and nonradiative (**4,5**) time constants for the samples obtained from OTAS.

| Sample | $\tau_{electron\ trapping}$ / ps | $\tau_{nonradiative}$ / ps |
|---|---|---|
| CIS100% | $1.18 \pm 0.03$ | $17 \pm 1.0$ |
| CIS20% | $18^{1} \pm 1$ | $550 \pm 30$ |
| CZIS200ºC | $85 \pm 8$ | $>1000$ |

[1]The electron trapping lifetime for CIS20% has been taken as an average of the two middle constants since the amplitudes are similar.

**Time-resolved X-ray absorption spectroscopy**

In order to gather more insights into the behavior of holes, we employed TR-XAS by probing the Cu K-edge. We chose to probe there because Cu contributes to a large degree to the valence band states, and the CHS is also attributed to Cu defects. The data was acquired at the SACLA XFEL facility.

The left side of Fig. 4 shows the steady-state and time-resolved XANES spectra at the Cu K-edge, and the right side shows the dynamics at specific energies. We can observe that the steady-state spectra, which in this case are obtained while the pumping laser at 520 nm is off, are similar to the ones obtained at ALBA synchrotron (Fig. 2A). This confirms that we are working with equivalent samples. We can observe in Figs. 4a and 4c that immediately after photoexcitation (1.2 ps, 0.14-0.54 ps, and 0.44 ps pump-probe delays, for CZIS200ºC, CIS100%, and CIS20%, respectively) the typical signature of $Cu^{2+}$ is present as a photoinduced absorption at 8978 eV and a large negative band denoting a blue-shift of the edge. According to our steady-state results, the ground state of CIS QDs is composed of $Cu^{+}$, but excitation at the bandgap of the material photo-induces oxidation from $Cu^+$ to $Cu^{2+}$. Therefore, this result agrees with the theoretical calculations showing that the valence band of CIS is mainly composed of Cu atomic orbitals[24] and with previous time-resolved synchrotron measurements.[29]

Looking at the time-resolved traces in Figs. 4B-D, we observe that the dynamics of the CZIS200ºC sample are flat in a time window of 1.5 ps, similar to what we have observed in the OTAS results. In contrast, the TR-XAS dynamics for the CIS100% and the CIS20% samples reveal ultrafast decay processes (Figs. 4D and 4F), due to a hole transfer to trap states unrelated to Cu (**2**). Moreover, for CIS100%, in less than 3 ps, the $Cu^{2+}$ suffers a reduction that not only brings it back to the initial $Cu^+$ state, but it is also mixed with what can be interpreted as structural rearrangement. The former is seen as a disappearance of the large negative band, while the latter is interpreted from the persistent GSB of the 8984 eV peak (1s→$4p_z$) and the appearance of positive side bands. This is explained as an increase in disorder due to thermal effects, producing a broadening.[46] We can reproduce the same transient by subtracting the steady-state signal of CIS20% to the one of CIS100% (Fig. S6.A), which means that the thermal energy of photoexcitation is augmenting the existing difference in crystallinity.

In summary, we observe a clear hole transfer away from Cu sites in both CIS100% and CIS20% samples. However, for CIS100% this comes together with an increased disorder, perhaps due to partial melting of the quantum dots. Meanwhile, CIS20% also shows the same delayed bleach of the 1s→$4p_z$ peak, but without the broadening due to disorder. Since neither sample is passivated, we are directly observing the hole transfer to a surface defect or a ligand (**2**). This occurs on the same timescales as the first relaxation steps seen in OTAS, which are associated with electrons (**3**), correlating with the heating effects.

Very similar behavior, as we detected for CIS100%, was observed for $Cu_2ZnSnS_4$ nanoparticles by Rein *et al.*[39] at an even higher fluence (40 mJ/cm$^2$ against our 13 mJ/cm$^2$). The observed phenomenon was also interpreted as an electron transfer to the Cu sites (hole transfer from them) mixed with structural rearrangement, where heating was influencing even more the measured signal. In the end, the delayed bleach of the 1s→$4p_z$ peak after hole transfer may originate from 3 different sources: i) the effect of electronic relaxation depositing thermal energy in the lattice, ii) the loss of coordination due to the oxidation of ligands, or iii) the effects of electrons in the conduction band. This last possibility was also pointed out by Rein *et al.* as the effect of promoting electrons between the antibonding orbitals of different atom pairs influences the structure in accordance with their EXAFS data. Further measurements would be needed to corroborate these hypotheses unambiguously. Alternatively, the fact that we observe ultrafast back-transfer in both non-passivated samples, while being absent in the passivated one, allows us to unequivocally assign the lack of PL signal in the former ones to sub-ps hole trapping at surface traps (**2**). These are probably related to the dodecane-thiol ligands, which have been observed to act as hole scavengers in other nanoparticles.[47] Thus, every other signal that is longer lived either in OTAS or TR-XAS should then be assigned to either the electrons in the conduction band, oxidized ligands or to thermal energy, since the valence band holes have been removed.

Clearly, our TR-XAS measurements of CZIS200ºC were limited by a poor signal-to-noise ratio since we performed those measurements at the lowest pump laser fluence at which they could be detected at SACLA XFEL. Thus, we are unable at this point to assign the few ps lifetime observed in OTAS to hole localization (**1**), as often suggested in the literature,[14,21,27] since we did not observe a compatible process in the Cu K-edge, even when probing at longer time-scale scans (Fig. S7). Nonetheless, we can rule out ultrafast trapping of photogenerated holes that hypothetically prevents Auger recombination processes, thus supporting our deduction from the steady-state XAS data that Cu-deficient QDs do not have a noticeable concentration of native holes in the form of $Cu^{2+}$. Moreover, it is plausible to assume that the localization of the hole is ultrafast and cannot be resolved, resulting in a direct formation of the clear $Cu^{2+}$ signal with no further evolution, as suggested by Berends *et al.*[17] Further X-ray experiments with higher sensitivity (at lower pump fluences to diminish the effect of laser-deposited heat inside the QD) must be carried out to confirm this hypothesis. Alternatively, we do prove that strong sub-ps processes in non-passivated samples are related to trapping and thermalization, since we observe the hole transfer and the heating effects on the structure.

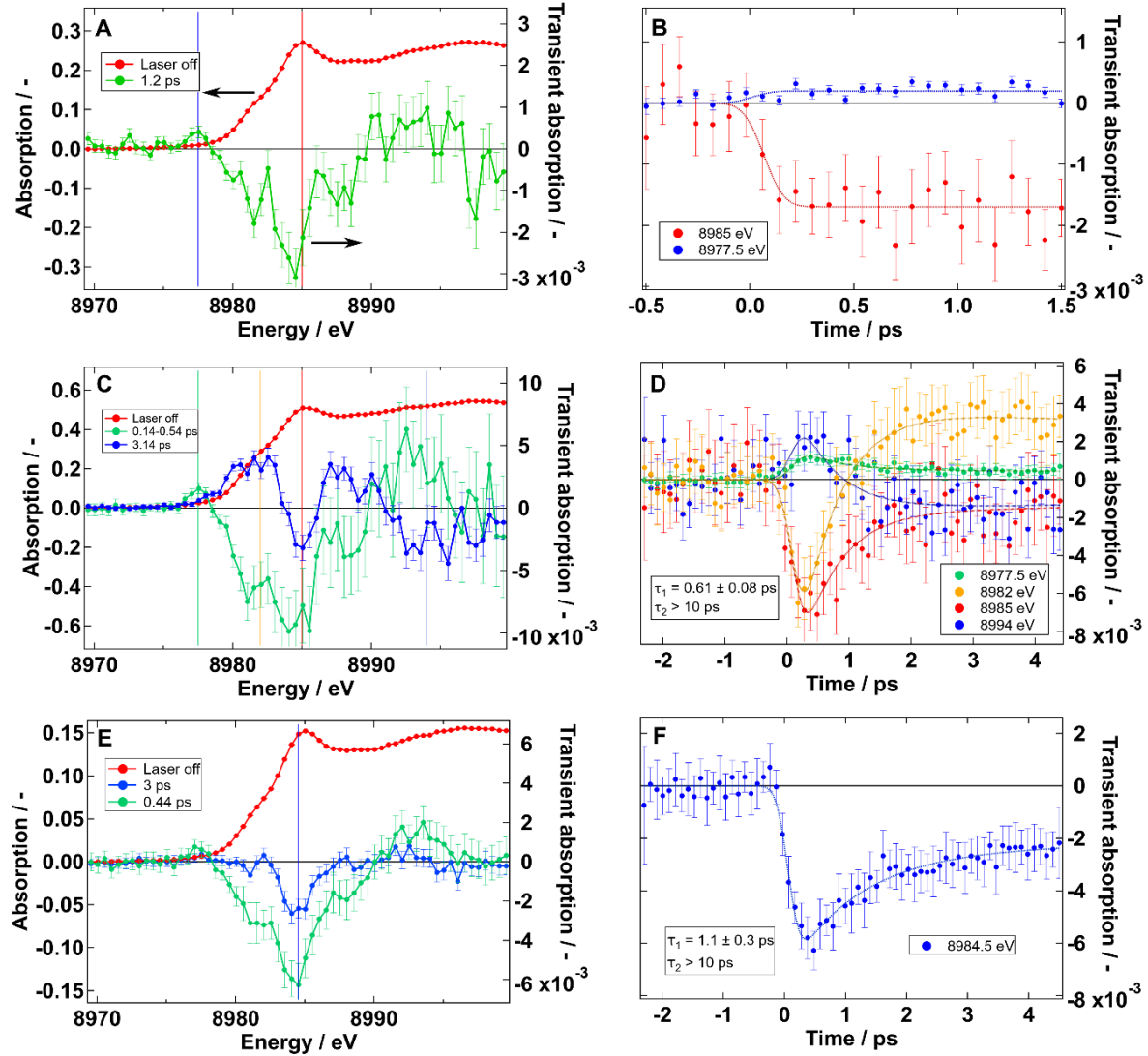

**Figure 4.** Steady-state (red line) and time-resolved XANES spectra at the Cu K-edge for CZIS200ºC (A), CIS100% (C), and CIS20% (E). TR-XAS dynamics at specific energies for CZIS200ºC (B), CIS100% (D), and CIS20% (F). The data were acquired by pumping at 520 nm with fluences of 3.3 mJ·cm$^{-2}$ (A,B), and 13 mJ·cm$^{-2}$ (C-F). The error bars correspond to the standard error.

**Fluorescence Up-conversion Spectroscopy**

Lastly, we performed fluorescence up-conversion spectroscopy (FLUPS) measurements of the non-stoichiometric samples, as a probe of emissive states. We find that the initial PL signal of CIS20% quickly decays with a lifetime of 1.0 ± 0.3 ps. This correlates well with the hole transfer time at the non-passivated surface (**2**), observed with TR-XAS (0.7 ± 0.2 ps). While both lifetimes are within their uncertainty, the slightly longer one for FLUPS can be explained through a combination of lower time resolution (~1 ps) and longer cooling time due to the 400 nm pumping. Interestingly, the PL is also described by a second lifetime of 30 ± 50 ps, this time with great uncertainty due to its weak amplitude. This second lifetime reveals the possibility of emissive recombination of trapped carriers, albeit with low probability. This can be produced from the trapped states, or through previous de-trapping. Due to the weak up-converted signal we cannot distinguish shifts of the PL band that could reveal its origin. Alternatively, the good passivation of the CZIS200ºC sample produces a comparatively stronger up-converted signal, with a much longer lifetime. The decrease in $\tau_1$ and $\tau_2$ compared to the OTAS results mainly originates from the use of 400 nm excitation. This should be seen as a red-shift of the emission in the first picoseconds due to the cooling to the band edge. However, our setup has a limited spectral region that can be measured. Consequently, the observed shift is very subtle (Fig. S11) and we mainly observe this effect as a decay of the signal. This decay can also be produced by a decreased radiative recombination rate due to hole localization (**1**).[27] Multi-excitonic effects due to the increase in fluence do not affect the fastest dynamics. As seen in Fig. S11, an almost 10-fold increase in fluence has small effects.

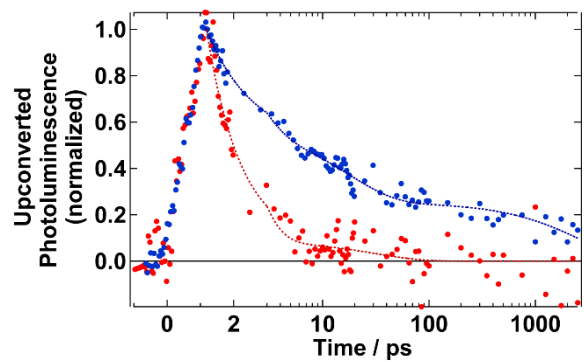

**Figure 5.** Wavelength-integrated FLUPS signal, for CZIS200ºC (blue) and CIS20% (red). Pumping at 400 nm, 3.7 mJ·cm$^{-2}$.

**Table 2:** Photoluminescence decay time-constants obtained from FLUPS.

| Sample | $A_1$ / cts | $A_2$ / cts | $A_3$ / cts | $\tau_1$ / ps | $\tau_2$ / ps | $\tau_3$ / ps |
|---|---|---|---|---|---|---|
| CZIS200 ºC | 0.6 ± 0.1 (45%) | 0.4 ± 0.1 (30%) | 0.33 ± 0.05 (25%) | 1.3 ± 0.8 | 20 ± 10 | 3·10$^3$ ± 2·10$^3$ |
| CIS20% | 0.25 ± 0.3 (93%) | 0.02 ± 0.02 (7%) | | 1.0 ± 0.3 | 30 ± 50 | |

According to our results, we propose a mechanism for the photophysical processes involved in CIS QDs. After photoexcitation, an electron is excited from the valence band to the conduction band leaving behind a hole. Since the valence band has a large proportion of Cu atomic orbitals, the hole is observed as an oxidation of Cu. Then, a hole is located at a Cu atom, forming the CHS. This localization appears to be ultrafast, or its effects are not observable with our signal-to-noise. $A_1$ in DAS is characterized by changes in the GSB resulting in a red-shift for all samples. This is produced by thermalization to the band edges, including fast trapping of charge carriers for CIS100% and CIS20% and structural rearrangement of the copper center. For the non-passivated samples, three processes occur:

thermalization; hole transfer, reducing back the $Cu^{2+}$ to $Cu^{+}$; and electron trapping. For the well-passivated samples, a partial electron trapping may occur which leads to $A_2$ in DAS or radiative recombination taking place by recombining the electron delocalized in the conduction band and the hole of $Cu^{2+}$. Due to the $3d^9$ electronic configuration of $Cu^{2+}$, the recombination with the electron in the conduction band is prompt. However, if the $Cu^{2+}$ is quickly reduced to $Cu^{+}$, the PL process will not occur. This explains why the PL efficiency for the stoichiometric sample, as well as the non-passivated non-stoichiometric one, is very low compared to the non-stoichiometric passivated one. In addition, the disorder produced in the stoichiometric sample introduces an additional electron trapping process which further quickens charge carrier recombination.

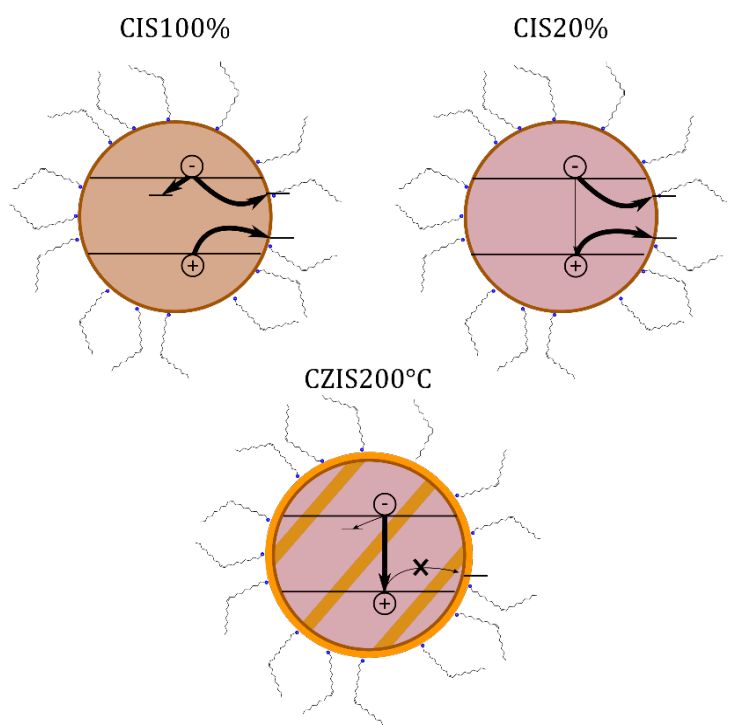

**Scheme 2**. Model describing the observed charge carrier dynamics in the three different samples, ignoring the CHS for simplicity. Thick arrows indicate the most probable processes in each system.

## Conclusions

We have employed a combination of X-ray and optical spectroscopy techniques, both static and time-resolved with femtosecond resolution, to shed new light into the photophysics and structure of CIS QDs. These include highly cutting-edge XFEL measurements with unique observables compared to previous measurements. We have shown that the improvement in PL efficiency with Cu-deficiency originates from a better crystallinity around Cu atoms. In addition, we have further studied the surface passivation with ZnS. Crystallinity has a stronger effect on electron trapping and thermalization-related structural effects, while surface passivation has a stronger effect on the trapping of both carriers at surface states. We observed at the Cu K-edge that the non-stoichiometric CIS QDs have a more ordered crystal structure than the stoichiometric analogs by looking at the broadening of features in XANES and the decreased coordination in EXAFS. The former was similarly simulated by introducing interstitial defects, while the latter also indicates a large number of vacancies and some degree of an amorphous phase. The steady-state XANES characterizations were also critical to understand that the ground state of CIS QDs is mainly composed of $Cu^+$, independently of the initial concentration of $Cu^+$ related to $In^{3+}$.

We have further studied our samples through time-resolved techniques probing both at the visible and the hard X-ray range. The latter allowed us to show for the first time the direct observation of hole transfer away from the QD core, with its Cu-dominated valence band, as a quenching mechanism in non-passivated samples. Furthermore, the hole transfer is produced within the same timescale as electron trapping, with its consequent heat deposition in the lattice and structural rearrangement. This rearrangement increases the structural disorder of the QDs. However, non-stoichiometric QDs with their improved crystallinity are resistant to this effect, even at high pump fluence conditions. The second lifetime is related to electron trapping processes in passivated samples, while it is already related to the recombination of trapped carriers in the other two samples.

These results have large implications in the understanding of the structural and compositional effects of synthesis stoichiometry, both contributing to the observed optical properties. Furthermore, they will help the development of CIS QDs and their applications by explaining the reason behind improved optical properties at Cu-deficient synthetic conditions.

## Methods

**CIS / Core Synthesis**

First, 0.0587 g of $InCl_3$ and 0.0238 g (1:1 Cu:In; 100%) or 0.004 g (0.2:1 Cu:In; 20%) of CuCl were weighed and placed in a 50 mL 3-neck flask. To this flask, 8 mL of 1-octadecene (ODE), 60 μL of oleic acid (OA), and 250 μL of dodecanethiol (DDT) were also added. The mixture was dried under vacuum at 90°C for 30 min. During this 30 min, a mixture with 0.038 g of sulfur (S) in 3 mL of oleylamine (OAm) was taken to ultrasound for 5 min. After this time, the solution of precursors in ODE was heated to 180 °C under an argon atmosphere for 5 min. Then, the temperature was lowered to 160 °C and 2 mL of the S-OAm solution were injected into the flask, monitoring for 10 min. The solution was cooled in an ice bath to room temperature (25 °C), under stirring and in an inert atmosphere (Argon).

After the synthesis step, the suspension was transferred to a Falcon centrifuge tube and 8.0 mL of isopropanol were added to purify the NCs. The tube was then taken to the centrifuge for 10 minutes at 7000 rpm. Finally, the supernatant was removed, and the nanoparticles were suspended in cyclohexane.

**CZIS / Core Synthesis**

First, 0.0587 g of $InCl_3$, 0.0274 g of $ZnCl_2$ and 0.004 g (0.2:1 Cu:In) of CuCl were weighed and placed in a 50 mL 3-neck flask. To this flask, 8 mL of 1-octadecene (ODE), 60 μL of oleic acid (OA) and 250 μL of dodecanethiol (DDT) were also added. The mixture was dried under vacuum at 90°C for 30 min. During this 30min, a mixture of 0.0257 g of sulfur (S) in 2 mL of oleylamine (OAm) was taken to ultrasound for 5 min. After this time, the solution in the 3-neck flask was heated to 180 °C under an argon atmosphere for 5 min. Then, the temperature was adjusted to 160 °C, and 2 mL of the S-OAm solution was injected into the flask, allowing it to react for 10 min. The solution was cooled in an ice bath to room temperature (25 °C), under stirring and in an inert atmosphere (Argon).

After the synthesis step, the suspension was transferred to a Falcon centrifuge tube and 8.0 mL of isopropanol were added to purify the NCs. The tube was then taken to the centrifuge for 10 minutes at 7000 rpm. Finally, the supernatant was removed, and the nanoparticles were suspended in cyclohexane.

**CZIS Core-Shell Synthesis:**

**Synthesis of Zn-OAm stock solution:**

First, 0.2725 g of $ZnCl_2$ were weighed and placed in a 50 mL 3-neck flask. To this flask, 4 ml of octadecene (ODE) and 1 ml of oleylamine (OAm) were also added. The mixture was dried under vacuum at 90 °C for 30 min. After this time, the solution was heated at 150 °C under an argon atmosphere for 10 min. Then, the temperature was adjusted to 50 °C.

**CZIS/ZnS Core-Shell Synthesis:**

The same procedure described above for the CZIS (item 2) was done, except for the purification step. With the pristine solution at room temperature and the Zn-OAm stock solution (item 3.1) at 50 °C, 5 mL of Zn-OAm solution were injected. Then, the system was heated to 200 °C and allowed to react for 30 min. Then, the solution was cooled in an ice bath to room temperature (25 °C), under stirring in an inert atmosphere.

After that, Isopropanol was added in a 1:1 ratio to the nanocrystal suspension and centrifuged at 9000 rpm for 10 minutes. The supernatant was discarded, and the tube remained open for ~5 minutes to dry isopropanol residues. The precipitate was suspended in cyclohexane.

**Transient absorption spectroscopy**

Transient absorption spectroscopy measurements were conducted using a Clark-MXR CPA-1 regenerative amplifier. The fundamental of the laser (775 nm, 1kHz, 120 fs, 1 mJ) was divided into two paths. One beam supplied a non-colinear optical parametric amplifier (NOPA) to generate 520 nm pulses, and filtered to the desired fluence to pump the sample. The second beam was sent through a $CaF_2$ crystal to generate a broadband supercontinuum by self-phase modulation spanning between 380 and 720 nm which was used as the probe. Due to technical circumstances, the CIS20% sample had to be probed with a supercontinuum generated with a sapphire crystal, limiting its bandwidth to 480-700 nm. A delay line was used to control the temporal delay between both pulses, which spatially overlapped on the sample. The probe pulse was divided before the sample position between a reference and a signal beams. The latter is sent through the sample, and both are collected into a prism spectrometer (Entwicklungsburo Stresing GmbH with a double CCD array. A home-made software recorded the normalized change in absorption (ΔA) in a shot-to-shot configuration. All measurements were performed at magic angle (54.7º) between the pump and probe to avoid anisotropy effects. The samples were measured in 2 mm thin quartz cuvettes with constant stirring with a magnetic bar perpendicular to the incident beam.

**Steady-state X-ray absorption spectroscopy**

The steady-state XAS spectra were obtained at the BL-22 CLÆSS Beamline from the ALBA synchrotron in Barcelona (Spain).[48] The X-ray beam was obtained from a multi-pole wiggler, and monochromatized with a double-crystal monochromator employing Si(111) crystals. The beam was focused down to a spot of 200x50 $\mu m^2$ at the sample position. The samples were contained in closed liquid cells with Kapton® windows, and the absorption was measured in total fluorescent yield mode. The result was checked for self-absorption by comparing it with the steady-state spectra obtained at SACLA, with a 100 μm jet, obtaining comparable spectra.

**Time-resolved X-ray absorption spectroscopy**

The TR-XAS experiments were carried out at the X-ray free electron laser facility SPring-8 Angstrom Compact free-electron LAser (SACLA) in Japan, at the BL3 beamline.[49,50] The FEL X-ray output of the undulator was set up with a central energy of 9 keV for Cu K-edge XANES, 500 μJ per pulse and a repetition rate of 30 Hz. At the sample position the X-rays were focused down to a spot of 2 μm in diameter. For excitation, a central wavelength of 520 nm was used, with a spot size of 1040 x 370 μm. The employed pulse energies ranged from 10 to 65 μJ, for fluences of 3.3 to 21 mJ cm$^{-2}$. The low repetition rate of SACLA XFEL poses a serious limitation on the signal-to-noise ratio of the recorded TR-XAS signal, leading finally to the necessity of using rather elevated optical pump laser fluences as described in the main text. We have confirmed experimentally the linearity of the detected TR-XAS signal at 13 mJ/cm$^2$ (Fig. S8) at the shortest time delays and used it throughout the present study. The synchronization system between pump and probe resulted in a jitter that fluctuated between 300 and 1200 fs. This was corrected on the presented data through the use of a timing tool.[51] The sample X-ray absorption was measured in a 100 μm cylindrical jet by capturing the X-ray fluorescence yield with a photodiode.

**Fluorescence up-conversion spectroscopy**

FLUPS measurements were carried out with a commercial fluorescence up-conversion spectrometer (HALCYONE, Ultrafast Systems). The fundamental output (100 fs and 1 mJ per pulse, λ = 800 nm, 1 kHz repetition rate) of a CPA Ti:sapphire laser (Spitfire, Spectra-Physics) was split into two beams. The first was attenuated and used as a gate. The beam went through a delay stage allowing us to resolve the fluorescence on a 3 ns time window. Then, it was focused onto a rotating BBO crystal for the up-conversion process. The second beam was sent through another BBO crystal for second harmonic generation (400 nm) and used as a pump. The beam was attenuated to the desired excitation pulse energy with a variable ND filter and set to magic angle polarization (54.7°) with respect to the gate using a half-wave plate. Then, the pump beam was sent onto the samples in a transmission geometry and a spot of 49 μm in diameter. The fluorescence was collected and focused into the up-conversion BBO crystal with parabolic mirrors. The up-converted beam was then focused and collected with an optical fiber, sent to a spectrograph, and the spectrum was measured with a thermoelectrically cooled CCD camera.

**Simulations**

Calculations of the edge region of the spectrum were performed using the finite difference method as implemented within the FDMNES package.[52] Throughout, we used a free form SCF potential of radius 6.0 Å around the absorbing atom. Broadening contributions due to the finite mean-free path of the photoelectron and to the core-hole lifetime were accounted for using an arctangent convolution.[53] Throughout, the crystal structures from Materials Project were used and adjusted as described in the main text.

# Supporting Information

Additional steady-state UV-vis characterization and PLQY, analysis of XANES and EXAFS data, additional OTAS spectra, analysis of TR-XAS spectra and supplementary plots, TRPL results, analysis of FLUPS data with complementary plots, further information on fitting models.

# Acknowledgments

AB is grateful to the Spanish "Ministerio de Universidades" and the "Plan de Recuperación, Transformación y Resiliencia", as well as the UAM, for his "Margarita Salas" grant (ref. CA1/RSUE/2021-00809). In addition, he receives funding from the European Union's Horizon 2020 research and innovation programme under the Marie Sklodowska-Curie agreement No. 101034431 and from the "Severo Ochoa" Programme for Centres of Excellence in R&D (CEX2020-001039S / AEI / 10.13039/501100011033. WG acknowledges funding from Spanish MIU through "Ayudas Beatriz Galindo" (BEAGAL18/00092), Comunidad de Madrid and Universidad Autónoma de Madrid through "Proyectos de I+D para Investigadores del Programa Beatriz Galindo" grant (Ref. SI2/PBG/2020-00003) and from Spanish MICIU through "Proyectos de I+D+i 2019" grant (Ref. PID2019-108678GB-I00). The steady-state XAS experiments were performed at the CLÆSS beamline at ALBA Synchrotron with the collaboration of ALBA staff, proposal 2021095311. The TR-XAS XFEL experiments were performed at the BL3 (EH2) of SACLA with the approval of the Japan Synchrotron Radiation Research Institute (JASRI), proposal 2021B8047. BRCV and LAP thank São Paulo Research Foundation, FAPESP, under grants 2018/15574-6 and 2022/06470-8. BRCV also thanks FAPESP for the postdoctoral scholarship under grants 2020/16077-6 and 2024/01722-4. AFVF thanks FAPESP under the grant 2023/10395-4. AFN acknowledges the support from the FAPESP (grant no. 2017/11986-5) and Shell and the strategic importance of the support given by ANP (Brazil's National Oil, Natural Gas, and Biofuels Agency). LB acknowledges support from the Spanish Ministry of Science and Innovation through grant PID2021-122839NB-I00. The facilities provided by the Center for Ultrafast Lasers (CLUR) of Universidad Complutense de Madrid (FLUPS) are gratefully acknowledged. JC-G acknowledges the MICINN-FEDER (No. PID2021-128313OB-I00), support from the Regional Government of Madrid (NMAT2D-CM), a Research Consolidation Grant (No. CNS2022-36191) and project PDC202-314587-1I00 from the Spanish Ministry of Science and Innovation. VVM acknowledges grants TED2021-131906A-100 and RYC2022-035200-I funded by Spanish Ministry of Science, Innovation and Universities (10.13039/501100011033) and support from the Regional Government of Madrid (2019-T2/IND-12737 and 2024-T1/TEC-31349). SGO is grateful to the Spanish Ministry of Science and Innovation for a Ph.D. grant (FPI, PRE2019-09345). The authors thank Reinhold Wannemacher and Luis Colmenar for their help with TRPL and PLQY measurements, respectively, at IMDEA Nanoscience.

# Supplementary Information

# Selective tracking of charge carrier dynamics in CuInS$_2$ quantum dots

Andrés Burgos-Caminal,*[1,2] Brener R. C. Vale,[3,4] André F. V. Fonseca,[5,4] Elisa P. P. Collet,[1,2] Juan F. Hidalgo,[1] Lázaro García,[2] Luke Watson,[6] Olivia Borrell-Grueiro,[7] María E. Corrales,[2,8] Tae-Kyu Choi,[9] Tetsuo Katayama,[10] Dongxiao Fan,[11] Víctor Vega-Mayoral,[1] Saül Garcia-Orrit,[1] Shunsuke Nozawa,[11] Thomas J. Penfold,[6] Juan Cabanillas-González,[1] Shin-Ichi Adachi,[11] Luis Bañares,[7,1] Ana F. Nogueira,[5] Lázaro A. Padilha,[3] Marco A. Schiavon,[4] and Wojciech Gawelda*[2,1,12]

1. Madrid Institute for Advanced Studies IMDEA Nanoscience, Ciudad Universitaria de Cantoblanco, Calle Faraday 9, 28049 Madrid, Spain.
2. Departamento de Química, Universidad Autónoma de Madrid, Ciudad Universitaria de Cantoblanco, Calle Francisco Tomás y Valiente 7, 28049 Madrid, Spain.
3. Instituto de Física Gleb Wataghin, Universidade Estadual de Campinas - UNICAMP, Campinas, São Paulo, Brazil.
4. Grupo de Pesquisa Química de Materiais, Departamento de Ciências Naturais, Universidade Federal de São João Del-Rei, Brazil.
5. Laboratório de Nanotecnologia e Energia Solar, Instituto de Química, Universidade Estadual de Campinas – UNICAMP, Campinas, São Paulo, Brazil.
6. Chemistry-School of Natural and Environmental Sciences, Newcastle University, Newcastle upon Tyne, NE1 7RU, UK.
7. Departamento de Química Física and Center for Ultrafast Lasers, Facultad de Ciencias Químicas, Universidad Complutense de Madrid, 2804 Madrid, Spain
8. Departamento de Química Física Aplicada, Universidad Autónoma de Madrid, Ciudad Universitaria de Cantoblanco, Calle Francisco Tomás y Valiente 7, 28049 Madrid, Spain.
9. XFEL Division, Pohang Accelerator Laboratory, POSTECH, Pohang, Gyeongbuk 37673, Republic of Korea.
10. Japan Synchrotron Radiation Research Institute, Kouto 1-1-1, Sayo, Hyogo 679-5198, Japan.
11. Institute of Materials Structure Science, High Energy Accelerator Research Organization (KEK), Japan.
12. Faculty of Physics, Adam Mickiewicz University, ul. Uniwersytetu Poznańskiego 2, 61-614 Poznań, Poland

Corresponding authors: wojciech.gawelda@uam.es, andres.burgos@imdea.org

# 1. Steady-state photoluminescence measurements

Steady-state photoluminescence (PL) measurements were recorded at IMDEA Nanoscience using a Horiba Fluorolog-3 fluorometer. The samples were measured in diluted conditions and the PL was captured at a perpendicular geometry. The samples were excited using 450 nm monochromatized light from a light bulb.

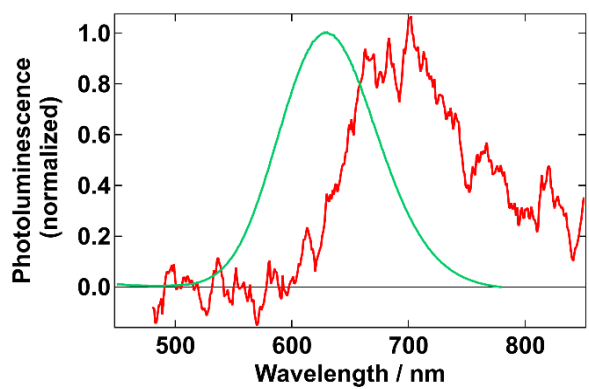

**Figure S1.** Normalized PL of CZIS200°C (green), and CIS20% (red) Only CZIS200°C is clearly emissive. CIS20% shows a very weak band, while CIS100% does not show any band above measurement artifacts and noise, and it is thus excluded.

Photoluminescence quantum yield (PLQY) measurements were carried out with a Quantaurus Plus, Hamamatsu, Pl quantum yield spectrometer. The statistics was obtained from different measurements at different wavelengths (Table S1). The results do not vary as a function of wavelength.

| $\lambda$ / nm | 400 | 410 | 420 | 430 | 440 | 450 | 460 | 470 | 480 | 490 |
|---|---|---|---|---|---|---|---|---|---|---|
| PLQY / % | 29.5 | 30.1 | 29.1 | 29.7 | 29.6 | 30.0 | 29.8 | 30.4 | 28.6 | 31.6 |

## 2. Analysis of steady-state XANES and EXAFS data (CLAESS beamline, ALBA)

To be able to compare with our samples presented in the article (Fig. 2) we measured two references, $Cu_2S$ and $CuS$, at ALBA synchrotron. These are examples of $Cu^+$ and $Cu^{2+}$, respectively. Sulfides were chosen as closer references to $CuInS_2$.

Unlike the QD samples presented in the main text, which were measured as dilutions and in total fluorescence yield mode, these references were measured in the form of pellets dispersed in cellulose and in transmission mode.

All the data was treated with the Athena program,[1] correcting the background and normalizing to the EXAFS region.

On Fig. S2 we analyze these reference spectra. On the left we can appreciate the small band characteristic of $Cu^{2+}$ at 8979 eV due to the $1s^2 \rightarrow 3d^9$ transition. This is easier to notice in the derivative spectrum (right) as shown by the arrow in the blue curve. However, our reference measurements present rather broad features, and substantial noise, unlike our sharper sample measurements in Fig. 2. This is a result of measuring the references in the form of pellets, in contrast to the sample liquid colloidal suspensions. Better examples can be found in the literature.[2,3]

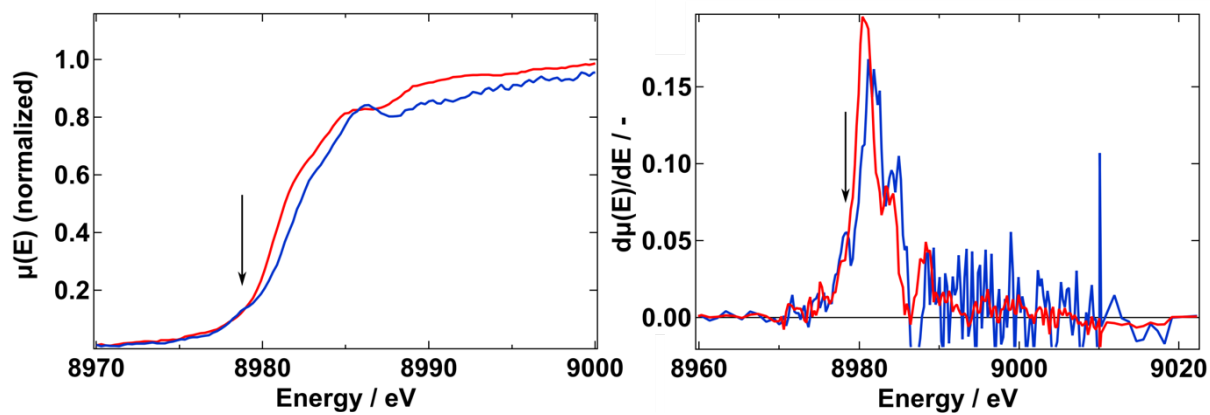

**Figure S2.** Reference XANES spectra (CuS, blue, and $Cu_2S$, red) at the Cu K-edge. Notice the signature of 1s-3d transition, as well as the general chemical shift of all features.

When compared with the XANES spectrum of CZIS200ºC in Fig. S3, all the pre-edge features align with the $Cu^+$ reference ($Cu_2S$).

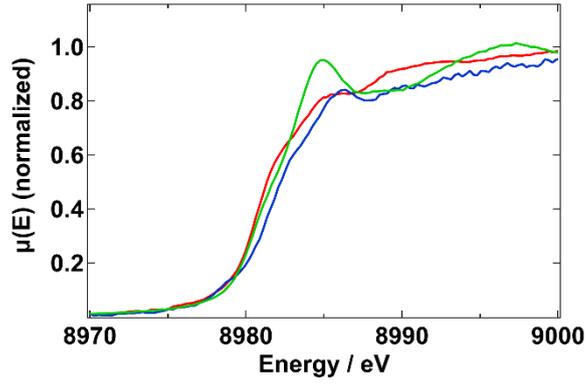

**Figure S3**. Comparison between the XANES spectra of the two references (CuS, blue, and $Cu_2S$, red) and CZIS200ºC (green).

EXAFS analysis was carried out with the Artemis program,[1] applying the in-built FEFF6 for the fit. All samples were fitted in a similar way.

k-range = 2.1 - 13
k-weight = 1,2
R-range = 1.35 – 5

The background removal was carried out with a k-weight of 2, a Rbkg of 1.3 and an E0 of 8983.3 eV.

The crystal structure of $CuInS_2$ was obtained from a reference and used for the FEFF calculation.[4] The resulting direct scattering paths with S (1st shell), Cu and In (2nd shell) were used to fit the EXAFS equation:[5]

$$\chi(k) = \sum_j \frac{S_0^2 N_j f_j(k) e^{-2R_j/\lambda(k)} e^{-2k^2\sigma_j^2}}{kR_j^2} \sin[2kR_j + \delta_j(k)], \qquad (S1)$$

Where we are interested in fitting $N_j$, the coordination number, $R_j$, the near neighbor distance, and $\sigma_j$, the mean square disorder. $S_0^2$ can be obtained from literature or from the fit of a well-known reference.

We obtained reasonable fits with R-factors of 0.0135, 0.0156 and 0.0232 for CIS100%, CIS20% and CZIS200ºC, respectively.

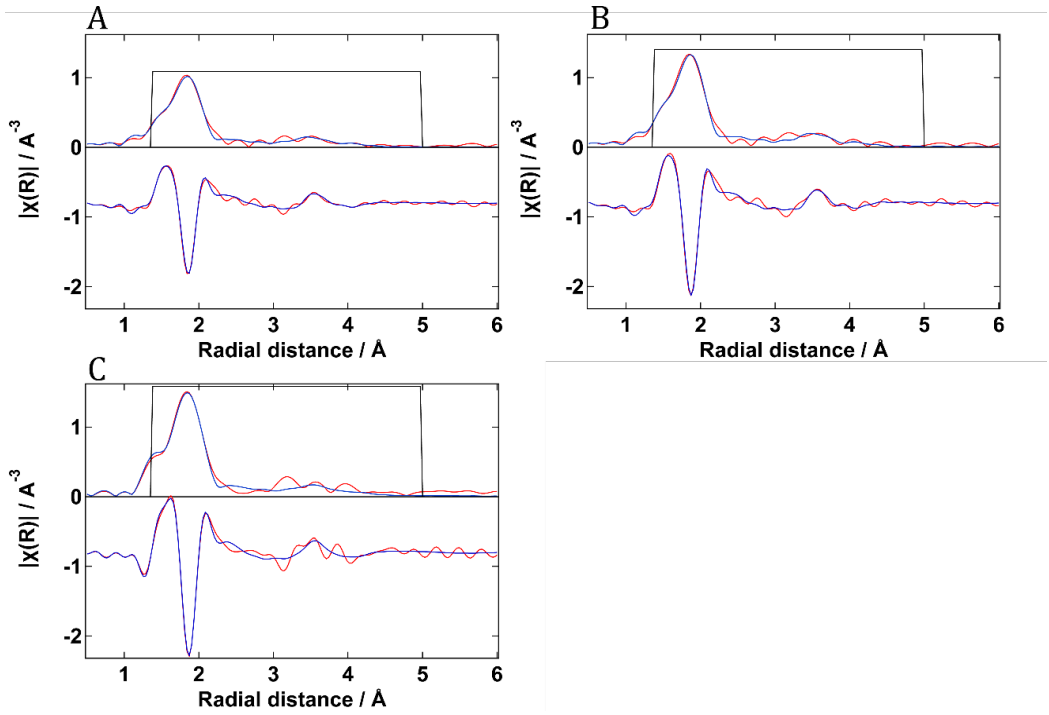

**Figure S4.** FEFF6 fits of CIS100% (A), CIS20% (B), and CZIS200°C (C).

**Table S2:** Cu K-edge EXAFS FEFF fitting parameters for the single scattering path with the first shell of S atoms. We assume an $S_0^2$ of 0.9 as a reasonable estimate, obtained from an EXAFS fit of metallic Cu.

| Sample | N·$S_0^2$ | N | $\sigma^2$ | R |
|---|---|---|---|---|
| CIS100% | 2.1 ± 0.1 | 2.3 ± 0.1 | 0.0076 ± 0.0009 | 2.301 ± 0.005 |
| CIS20% | 2.7 ± 0.2 | 3.0 ± 0.2 | 0.008 ± 0.001 | 2.316 ± 0.005 |
| CZIS200°C | 3.0 ± 0.3 | 3.4 ± 0.3 | 0.007 ± 0.001 | 2.314 ± 0.007 |

## 3. Time-resolved optical TAS data

In Fig. 2 we show the result of a global fit on the OTAS data. More details are given in section 8. Instead, here in Fig. S5 we show the ΔA spectra at given times.

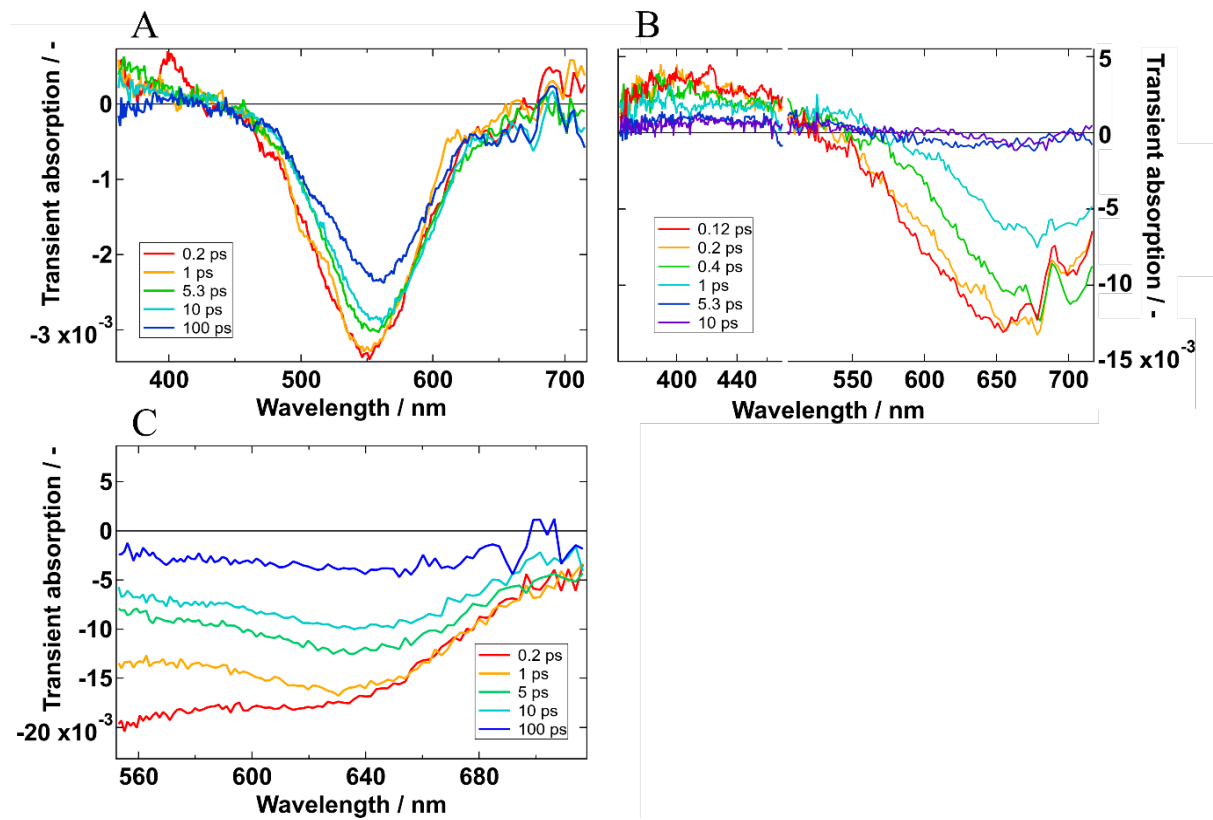

**Figure S5.** Transient absorption spectra at different time delays of CZIS200˚ C(A), CIS 100% (B), and CIS20% (C).

## 4. Analysis of time-resolved XANES data (SACLA XFEL)

To further illustrate the processes assigned to the transients of our TR-XAS data, we have simulated three transients.

First, the difference between the XAS spectra of CIS100% and CIS20%, to show that the increase in disorder between these two samples is very similar to the obtained transient in the former.

Second, the effect of a general broadening due to disorder. This is obtained by convoluting the XAS spectrum of CIS100% with a Gaussian profile. The result is similar to the long-lived transient in CIS100%. However, the transient lacks the pronounced peak at 8979.5 eV. This is because in the data the broadening affects mainly the 1s→4$p_z$ peak at 8985.

Third, the effect of a general blue-shift of the spectrum of CIS100% due to oxidation. The broad negative feature is similar to the early times transient. However, it lacks the appearance of the $1s^2$→$3d^9$ peak at 8977.5 eV due to the formation of $Cu^{2+}$, present in the experimental data.

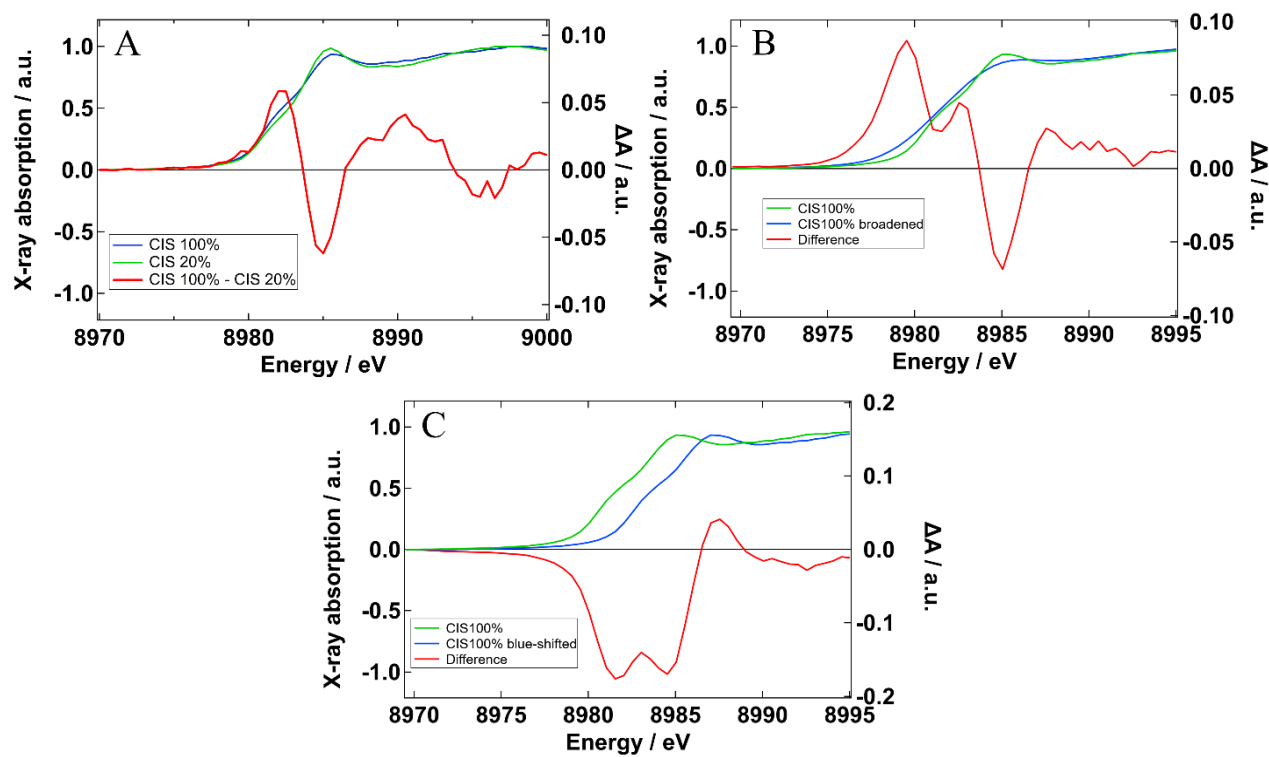

**Figure S6**. A) Difference XAS between CIS100% and CIS20% showing a very similar change to the long-lived transient of CIS100%. B) Simulated effect of a general broadening on the XAS spectrum of CIS100%. The shape resembles closely the transient at longer times, with an additional band at low energy. C) Simulated effect of a general blue-shift (oxidation) on the XAS spectrum of CIS100%. The large negative band agrees with the transient spectrum at early times, although it lacks the appearance of the $1s^2$→$3d^9$ peak at 8977.5 eV.

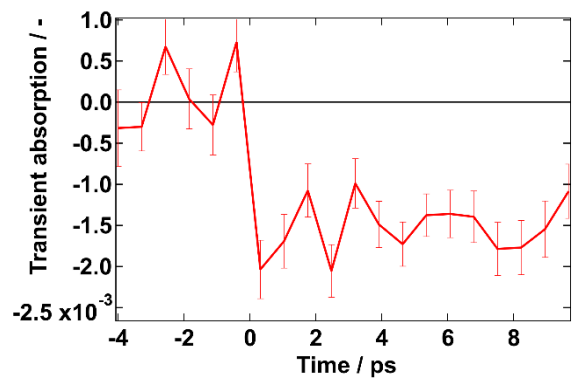

**Figure S7**. Extended window for the time-resolved XANES of CZIS200ºC at 8985.5 eV, shown in Figure 4.

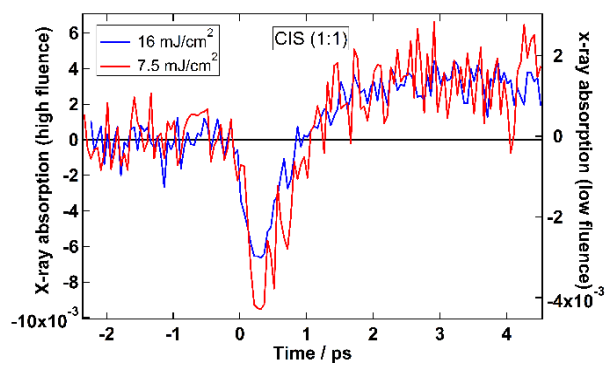

**Figure S8**. Fluence dependent TR-XAS, where the signal scales linearly with the fluence and the decay rate are similar within the noise.

## 5. Time-resolved photoluminescence measurements

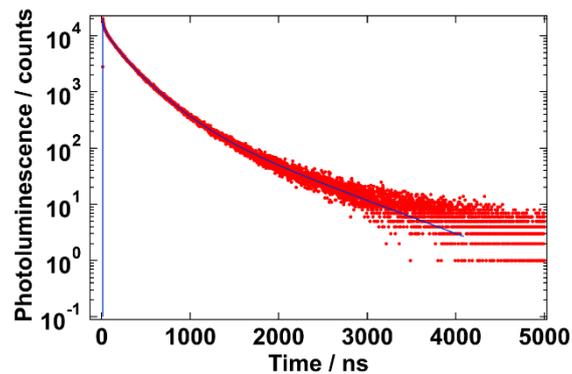

**Figure S9.** Time-resolved PL of CZIS200°C showing a multiexponential decay.

| | |
|---|---|
| IRF/ns | 0.996 ± 0.005 |
| t0/ns | 11.475 ± 0.003 |
| A1/cts | 4.34·10³ ± 7·10¹ |
| A2/cts | 3.91·10³ ± 6·10¹ |
| A3/cts | 7.3·10³ ± 3·10² |
| A4/cts | 5.6·10³ ± 3·10² |
| A5/cts | 6.5·10² ± 9·10¹ |
| τ1/ns | 3.5 ± 0.1 |
| τ2/ns | 19.2 ± 0.5 |
| τ3/ns | 2.73·10² ± 8 |
| τ4/ns | 1.2·10² ± 4 |
| τ5/ns | 7.4·10² ± 3·10¹ |

**Table S3.** Fit results of the TRPL data in Figure S10 for a 5-exponential model.

We can assume that the last three components will be the ones corresponding to particles free of electron trapping. Then the averaged radiative lifetime is calculated as an intensity-weighted average of the last 3 components as follows:

$$\frac{\sum_n \tau_n^2 A_n}{\sum_n \tau_n A_n} \tag{S2}$$

Obtaining an average radiative lifetime of 310 ns.

## 6. Analysis of FLUPS data

The time-resolved detection of the PL through up-conversion depends highly on the PL intensity at any given time. If a material has a very long PL lifetime, it will emit few photons per 100 femtoseconds. Therefore, even if the material has high PL quantum yield (PLQY) it will be hard to detect them with FLUPS. This is the case for CIS QDs, even for CZIS200ºC, hence the use of considerable fluences in the measurements.

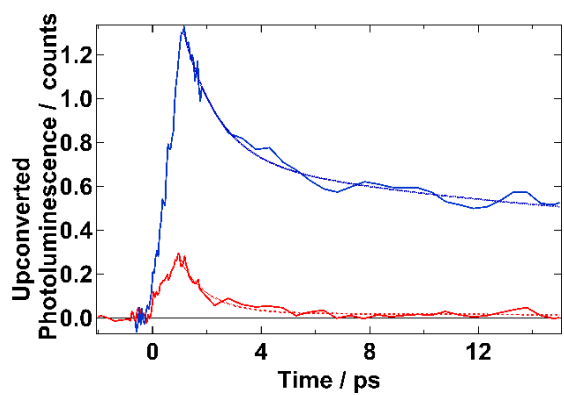

**Figure S10.** Linear scale and not normalized version of Fig 5, showing CZIS200°C (blue) and CIS20% (red). The time zero has been shifted in Fig. 5 to accommodate the log scale.

Another source of difficulty in FLUPS is the fact that the response of the setup is not equal at all wavelengths. A photometric correction has been employed in the past to correct this problem.[6] However, our poor spectral resolution compared to a standard UV-vis makes the spectra hardly comparable. Furthermore, with CIS QDs there is a larger issue. To avoid contamination from the 400 nm light employed as pump, a short-pass filter is employed that only lets through the up-converted UV radiation. This translates in an abrupt cut of PL detection at 650 nm. In conclusion, the obtained spectra are limited, and these limitations should be considered. Here, we show a qualitative comparison in Fig. S11, while throughout the other figures a spectral integration is carried out to maximize the signal of the temporal traces. This seems to distort slightly the rise of the signal. Thus, instead of employing the convoluted exponential explained in the next section, we fitted standard exponential equations to the decays starting at the maximum.

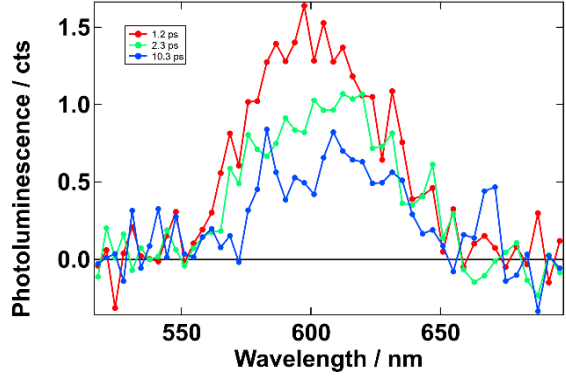

**Figure S11.** FLUPS spectra of CZIS200ºC at different time delays. Notice that a shift on the maximum is apparent. The detection efficiency of our setup quickly decays above 600 nm, cutting completely at 650 nm. Consequently, the shift is only observed as a faster decay of the blue side of the band.

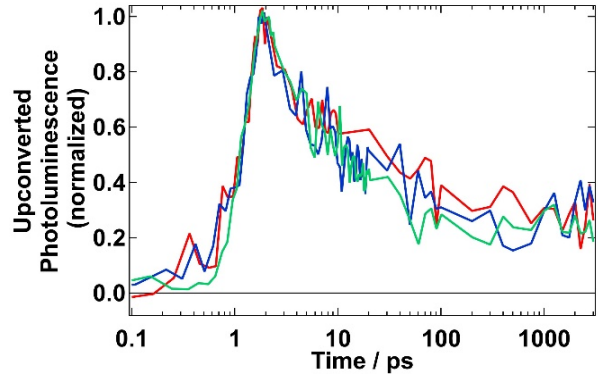

**Figure S12.** Upconverted photoluminescence of CZIS200°C at 0.8 (red), 2.6 (blue), and 5.8 (green) mJ/cm$^2$.

## 7. Exponential models

To carry out the fits of the time-resolved dynamics we employed an exponential decay convoluted with a gaussian resulting in the equation:

$$S(t) = A\frac{1}{2}\exp\left(\frac{4ln(2)w^2}{\tau^2} - \frac{t-t_0}{\tau}\right)\left(1 + \text{erf}\left(\frac{t-t_0}{4\sqrt{ln(2)}\,w} - \frac{2\sqrt{ln(2)}\,w}{\tau}\right)\right) \quad (S3)$$

Where A is the amplitude, w is the Full width at half maximum, τ is the lifetime, and t0 is the time zero. For a multi-exponential decay, we employ the sum of several such equations.

**8. Global fit**

We carried out global fits for the OTAS and TR-XAS data using the Igor Pro built-in in package. For any global fit, a set of equivalent multiexponential equations (sum of Eq. S3) equal to the number of time traces to be analyzed is fitted simultaneously. Each time trace corresponds to a different wavelength or energy. The parameters $w$ and $\tau_n$ are treated as common for all time traces, obtaining best fit values for all of them. On contrary, a different value of $A_n$ parameter is fitted for each time trace. The spectral representation of $A_n$ against wavelength or energy is called decay associated spectrum (DAS), as shown in Fig. 3. The DAS allow us to interpret the spectral changes produced by each exponential decay with their associated $\tau$.